\documentclass[a4paper]{article}
\usepackage{amsfonts,amssymb,amsmath,amsthm,cite}
\usepackage{ucs} 
\usepackage[utf8x]{inputenc}
\usepackage{graphicx}
\usepackage[english]{babel}
\usepackage{slashed}
\usepackage{textcomp}
\usepackage{bm}
\usepackage{titlesec}
\usepackage{tikz-cd}
\usepackage{adjustbox}
\usepackage{multirow}
\usepackage{etoolbox}
\patchcmd{\thebibliography}{\section*}{\section}{}{}
\usepackage{mathtools}
\evensidemargin=\oddsidemargin
\begin{document}
\vspace{10mm}
\begin{center}
	\large{\textbf{Summation of power singularities}}
\end{center}
\vspace{2mm}
\begin{center}
	\large{\textbf{A. V. Ivanov}}
\end{center}
\begin{center}
St. Petersburg Department
of Steklov Mathematical Institute
of Russian Academy of Sciences,\\
27 Fontanka, St. Petersburg, 191023, Russia
\end{center}

\begin{center}
E-mail: regul1@mail.ru
\end{center}

\vspace{10mm}

\textbf{Abstract.} In this paper, we investigate an example of summation of non-logarithmic singularities of a specific type in a two-dimensional non-linear sigma model. As a result of the study, we obtained an explicit formula, which, upon formal expansion in terms of the coupling constant, reproduces a particular part of the quantum action. Additionally, the paper introduces a new auxiliary function and discusses some of its properties.

\vspace{2mm}

\newpage

\section{Introduction}

The perturbative methods of quantum field theory \cite{3,9,10} are often associated with concepts such as renormalization, regularization, and singularity, see \cite{6,7,105}, since the coefficients of formal series can contain divergent quantities. Usually, the study of such objects is carried out in a sequential manner, as higher correction contributions depend on the process of renormalizing the previous ones.

In some particular cases, a special type of singularity can be selected, which is present in all correction terms and has predictable behavior. The main logarithmic singularities are usually considered as such objects. However, such cases are rare. See, for example, \cite{36-2,36-3,36-4,36-5} or, moving on to more abstract considerations, calculations in \cite{29-1-0,29-1-1,o-3,o-4,o-5,o-6,o-7}.

In this paper, we consider an example of summing the main non-logarithmic (power) singularities in the two-dimensional principal chiral field model, see \cite{HB3,36-6,HB,36-7,q-12,sig1,q-8,q-10}. Smoothing of the $\delta$-functional is used as a regularization method, which can be implemented using a cutoff regularization \cite{nkf-1,Sh,w6,w7,Khar-2020}, in particular, by the cutoff in the coordinate representation \cite{Ivanov-2022,Iv-2024,sk-b-20,sksk}. The presented study is a continuation of the series of papers \cite{Iv-244,Ivanov-Akac,sk-b-19,AIK-25} devoted to the study of the structure of singularities in non-linear sigma models.

The text is organized into three main sections: the problem statement, the calculation process, and proofs of certain properties. It concludes with a discussion of comments and open questions.

\section{Problem statement}
Consider the space $\mathbb{R}^2$ and the Lie algebra $\mathfrak{g}$ of the group $\mathrm{SU}(n)$, where $2\leqslant n\in\mathbb{N}$, see \cite{2}. Generators $t^a$, where $a\in\{1,\ldots,n^2-1\}$, of the algebra $\mathfrak{g}$ satisfy the standard relation $[t^a,t^b]=f^{abc}t^c$, where $f^{abc}$ are totally antisymmetric real structure constants. The latter also have the property $f^{abe}f^{abc}=n\delta^{ec }$, see Appendix C in \cite{13}. Next, the symbol "$\mathrm{tr}$" denotes the trace over group indices.

In a recent paper devoted to the study of three-loop singularities in a two-dimensional principal chiral field model \cite{AIK-25}, an ansatz for a renormalized quantum action was written out as follows
\begin{equation}\label{36-1}
W_{\mathrm{ren}}[B,\Lambda]=
\frac{S[B]}{4\gamma^2_0}-\frac{1}{2}\big(\ln\det(G_{\mathrm{reg}})-\varkappa_0\big)-\bigg[\mathbb{H}_0^{\mathrm{sc}}\exp\bigg(
\sum_{k=3}^{+\infty}\frac{\gamma^{k-2}_0}{k!2}\Big(\Gamma_{k}+\Gamma_{r,k-2}\Big)\bigg)
-\sum_{k=1}^{+\infty}\gamma^{2k}_0\varkappa_k\bigg].
\end{equation}
Here, $S[B]$ denotes the classical action that depends on a background field, which, in turn, satisfies the quantum equation of motion, see \cite{Ba-2}. The function $G_{\mathrm{reg}}^{ab}(\cdot\,,\cdot)$ is a regularized Green's function with the regularization parameter $\Lambda$, and the value $\gamma_0=\gamma_0(\gamma,\Lambda)=\gamma+\ldots$ stands for the renormalized coupling constant. Further, for $k\geqslant1$, the representation is valid for vertices with $k$ external lines
\begin{equation}\label{36-2}
\Gamma_{k+2}[\phi]=-\int_{\mathbb{R}^2}\mathrm{d}^2x\,
\big(\partial_{x_\mu}\phi^{a_1}(x)\big)\big(\phi^k(x)\big)^{a_1a_{k+1}}
D_\mu^{a_{k+1}b}(x)\phi^b(x),
\end{equation}
where $D_\mu^{ab}(x)=\delta^{ab}\partial_{x^\mu}-f^{acb}B_\mu^{c}(x)$ denotes the covariant derivative, and notation has also been used for the fluctuation field $\phi^{ab}(x)=f^{acb}\phi^{c}(x)$ and their products
\begin{equation*}
\big(\phi^k(x)\big)^{a_1a_{k+1}}=\prod_{i=1}^{k}\phi^{a_ia_{i+1}}(x).
\end{equation*}
In turn, $\Gamma_{r,k-2}$ denotes a counter vertice, which is a finite linear combination of auxiliary vertices with coefficients singular in the parameter $\Lambda$ and the number of external lines not exceeding the number of $k-2$. Next, the values of $\varkappa_k$ subtract divergences with densities independent of the background field, and the operator $\mathbb{H}_0^{\mathrm{sc}}$ connects the outer lines of the vertices in all possible ways using the deformed Green's function $G_\Lambda^{ab}(\cdot\,,\cdot)$ and leaves only the strongly connected vacuum part. Note that $G_\Lambda$ is uniquely constructed using the function $G_{\mathrm{reg}}$ and coincides in the main order with the deformed function $2G^{\Lambda}_0(x-y)\delta^{ab}$ for the free Laplace operator $A(x)=-\partial_{x_1}^2-\partial_{x_2}^2$. The latter is conveniently understood in the sense of the Fourier transform
\begin{equation}\label{36-13}
A(x)G^{\Lambda}_0(x)=\int_{\mathbb{R}^2}\frac{\mathrm{d}^2k}{4\pi^2}\,e^{ikx}\rho(|k|/\Lambda),
\end{equation}
where $0\leqslant\rho(\cdot)\leqslant1$ is called a regularizing (smoothing) function. Due to the property $\rho(0)=1$, when removing the regularization $\Lambda\to+\infty$, the right-hand side of the equality tends to the $\delta$-functional in the sense of generalized functions, see \cite{Gelfand-1964,Vladimirov-2002}.

In the process of studying two and three loops, it was noticed that, in the $n$-loop coefficient of quantum action, it becomes necessary to introduce an auxiliary vertex of the form
\begin{equation}\label{36-7}
\mathrm{V}_{2n-2}[\phi]=\int_{\mathbb{R}^2}\mathrm{d}^2x\,\mathrm{tr}\big(\phi^{2n-2}(x)\big)
\end{equation}
with $2n-2$ external lines. In this case, such a vertex must enter with a coefficient proportional to $\Lambda^2\gamma^{2n-2}$ and some function $\rho_{2n-2}$, which is constructed exclusively from combinations of deformed $\delta$-functionals $A(x)G^{1}_0(x)$. We call such vertices the main ones. It is convenient to fix the coefficient so that $\Gamma_{r,2n-2}=2(2n!)\Lambda^2\rho_{2n-2}\mathrm{V}_{2n-2}+\ldots$ Note that in the parts of the effective action with the number of loops more than "$n$", vertices $\mathrm{V}_{2n-2}$ may also appear, but they are not the main ones, since they enter with a higher degree of the coupling constant and/or with a coefficient that is constructed not only using $A(x)G^{1}_0(x)$, but also functions containing a different number of derivatives of $G^{\Lambda}_0(x)$. \\

\noindent\textbf{The main aim.} Find the sum of all the main auxiliary vertices that occur in the quantum action \eqref{36-1}, that is, calculate the value of the form
\begin{equation*}
\Psi=\Lambda^2
	\sum_{n=2}^{+\infty}\gamma^{2n-2}\rho_{2n-2}\mathrm{V}_{2n-2}.
\end{equation*}
The answer is given by formula \eqref{36-19}.

\section{Calculation}
Note that some simplifications can be used to find the main vertices. Indeed, for reasons of dimensionality conservation, it follows that $B=0$ can be chosen. In this case, all the odd vertices from \eqref{36-2} vanish, while the even ones take the form
\begin{equation*}
	\Gamma_{0,2k+2}[\phi]=-\int_{\mathbb{R}^2}\mathrm{d}^2x\,
	\big(\partial_{x_\mu}\phi^{a_1}(x)\big)\big(\phi^{2k}(x)\big)^{a_1a_{2k+1}}
	\partial_{x^\mu}\phi^{a_{2k+1}}(x).
\end{equation*}
Then, given the explicit form of quantum action \eqref{36-1}, the desired combination $\Psi$ can be represented as the functional
\begin{equation}\label{36-8}
-\mathbb{\dot{H}}_0^{\mathrm{sc}}\exp\bigg(
\sum_{n=1}^{+\infty}\frac{\gamma^{2n}}{(2n+2)!2}\Gamma_{0,2n+2}\bigg),
\end{equation}
where the operator $\mathbb{\dot{H}}_0^{\mathrm{sc}}$ arbitrarily connects all fields with derivatives, then leaving only the main part of the asymptotics with respect to the parameter $\Lambda$ for the strongly connected part. It is proportional to $\Lambda^2$. To analyze the last combination, consider an arbitrary finite monomial. To do this, we define a multiindex $\alpha=\{\alpha_i\}_{i=1}^{+\infty}$, an arbitrary element $\alpha_k\in\mathbb{N}\cup\{0\}$ of which denotes the degree of the vertex $\Gamma_{0,2k+2}^{\alpha_k}$, as well as two numbers
\begin{equation*}
\dot{\alpha}=\sum_{k=1}^{+\infty}\alpha_k,\,\,\,\hat{\alpha}=\sum_{k=1}^{+\infty}k\alpha_k.
\end{equation*}
Then the value of \eqref{36-8} is the sum over all possible nonzero multiindices $\alpha$ with combinations of the form
\begin{equation}\label{36-10}
-
\mathbb{\dot{H}}_0^{\mathrm{sc}}\prod_{n=1}^{+\infty}
\frac{\gamma^{2n\alpha_n}2^{-\alpha_n}}{[(2n+2)!]^{\alpha_n}\alpha_n!}
\Gamma_{0,2n+2}^{\alpha_n}.
\end{equation}
Let us calculate the application of the operator. To do this, we note that all possible combinations of fields with derivatives lead to diagrams of the "chain" type. Thus, preserving only the main part of the asymptotics and taking into account a symmetry coefficient, we obtain
\begin{equation}\label{36-11}
\mathbb{\dot{H}}_0^{\mathrm{sc}}\prod_{n=1}^{+\infty}
\Gamma_{0,2n+2}^{\alpha_n}=\Lambda^2
\mathrm{V}_{2\hat{\alpha}}\vartheta(\dot{\alpha})\frac{(-4)^{\dot{\alpha}}\dot{\alpha}!}{2\prod_{n\geqslant1}\alpha_n!},
\end{equation}
where the auxiliary function $\vartheta(\cdot)$ is defined by the equality
\begin{equation*}
\vartheta(n)=
\underbrace{AG^{1}_0\star\ldots\star AG^{1}_0}_\text{n\,\mbox{times}}\,(0)
\end{equation*}
for all $n\geqslant1$. The operation $\star$ denotes the convolution. Next, using representation \eqref{36-13} and a combination of convolution with the Fourier transform, the formula is rewritten as
\begin{equation*}
	\vartheta(n)=\int_{\mathbb{R}^2}\frac{\mathrm{d}^2k}{4\pi^2}\,\rho^n(|k|).
\end{equation*}
Thus, after substituting \eqref{36-11} into \eqref{36-10}, we get
\begin{equation*}
	-\frac{\Lambda^2}{2}\sum_{\alpha:\,\dot{\alpha}>0}\bigg(
	(-4)^{\dot{\alpha}}\dot{\alpha}!\mathrm{V}_{2\hat{\alpha}}\vartheta(\dot{\alpha})
	\prod_{n=1}^{+\infty}
	\frac{\gamma^{2n\alpha_n}2^{-\alpha_n}}{[(2n+2)!]^{\alpha_n}\alpha_n!\alpha_n!}\bigg).
\end{equation*}
Next, note that we can proceed to summation by components of the multiindex. To do this, we first use the definition for the vertex from \eqref{36-7} and the representation for the factorial
\begin{equation*}
	n!=\int_{\mathbb{R}_+}\mathrm{d}s\,s^{n}e^{-s}.
\end{equation*}
Then, by adding and subtracting the summand corresponding to $\dot{\alpha}=0$, we obtain a factorization of the sums, in each of which we can use the representation for the Bessel function of the first kind, see Section 8.441 in \cite{GR},
\begin{equation*}
\sum_{k=0}^{+\infty}\frac{(-s)^k}{k!k!}=J_0(2\sqrt{s}).
\end{equation*}
As a result, we obtain for $\Psi$ the following formula
\begin{equation}\label{36-16}
	-\frac{\Lambda^2}{2}\mathrm{Tr}_{x,k}
	\Big[\mathrm{Kh}\Big(\phi(x)\gamma,2\sqrt{\rho(|k|)}\Big)-\mathbf{1}\Big],
\end{equation}
where "$\mathbf{1}$" denotes the identity matrix, the integral operator $\mathrm{Tr}_{x,k}$ is given by the formula
\begin{equation*}
\mathrm{Tr}_{x,k}=\int_{\mathbb{R}^2}\mathrm{d}^2x
\int_{\mathbb{R}^2}\frac{\mathrm{d}^2k}{4\pi^2}\,\mathrm{tr},
\end{equation*}
and the auxiliary function $\mathrm{Kh}(\cdot\,,\cdot)$ is defined by the equalities
\begin{equation}\label{36-18-1}
	\mathrm{Kh}(u,v)=\lim_{j\to+\infty}\mathrm{Kh}_j(u,v),
\end{equation}
\begin{equation}\label{36-18}
	\mathrm{Kh}_j(u,v)=
	\int_{\mathbb{R}_+}\mathrm{d}s\,e^{-s}\Bigg[
	\prod_{n=1}^{j}J_0\bigg(\frac{u^nv\sqrt{s}}{\sqrt{n+1}}\bigg)\Bigg].
\end{equation}
Note that formula \eqref{36-16} can be reduced to the sum over the roots of characteristic polynomial of the matrix $\phi(x)$. Indeed, note that the matrix $\phi(x)$ is real and skew-symmetric. Therefore, see Chapter $\mathrm{VI}$ in \cite{36-1}, each root of its characteristic polynomial for each fixed "$x$" is either purely imaginary or zero. Moreover, imaginary ones come in pairs (with their conjugates). Notate the roots as $i\lambda^a(x)$, where $a\in\{1,\ldots,n^2-1\}$ and $\lambda^a(x)\in\mathbb{R}$. Then formula \eqref{36-16} is rewritten as
\begin{equation}\label{36-19}
\Psi=-\frac{\Lambda^2}{2}
\sum_{a=1}^{n^2-1}
\mathrm{Tr}_{x,k}
	\Big[\mathrm{Kh}\Big(i\lambda^a(x)\gamma,2\sqrt{\rho(|k|)}\Big)-\mathbf{1}\Big].
\end{equation}

\section{Properties}
Let us pay attention to some properties of the latter function, starting with the case of real arguments. First, it follows from the property $J_0(0)=1$ that $\mathrm{Kh}(u,v)=1$ for $u=0$ and/or $v=0$. Secondly, given the parity of the Bessel function, we obtain parity with respect to both arguments $\mathrm{Kh}(u,v)=\mathrm{Kh}(-u,v)=\mathrm{Kh}(u,-v)$. Next, assume that $v\neq0$ and $|u|>1$, and note that in this case the function $|u|^n/\sqrt{n+1}$ is increasing for all $n>N_1$ for some fixed $N_1\in\mathbb{N}$. Therefore, there exists a number $0\leqslant\delta_1<1$ such that
\begin{equation*}
\bigg|J_0\bigg(\frac{u^nv\sqrt{s}}{\sqrt{n+1}}\bigg)\bigg|<\delta_1
\end{equation*}
for all $n>N_1$. The number $\delta_1$ may depend on $v\sqrt{s}$. Therefore, the product in \eqref{36-18} contains arbitrarily many multipliers of $\delta_1$ in the limit. Considering that the rest of the product is limited to one, we get $\mathrm{Kh}(u,v)=0$ for all $v\neq0$ and $|u|>1$. Note that in this case, the integral and the limit transition can be permuted due to the presence of uniform convergence, see Chapter 14 in \cite{34-f-1}. Moreover, the latter relation can be strengthened. Let $|u|=1$ and $a=v^2s/4$, then there exists such a $N_2\in\mathbb{N}$ that for all $n>N_2$ we have $a<n+1$ and the estimate is valid
\begin{equation*}
\ln\bigg|J_0\bigg(\frac{u^nv\sqrt{s}}{\sqrt{n+1}}\bigg)\bigg|
\leqslant-\frac{a}{n+1}+\frac{a^2}{4(n+1)^2},
\end{equation*}
where we used the relation $\ln(1-s)\leqslant-s$ for $s\in[0,1]$. Thus, the absolute value of the product from \eqref{36-18} is estimated from above by
\begin{equation*}
\exp\Bigg(\sum_{n=N_2+1}^{+\infty}\bigg[-\frac{a}{n+1}+\frac{a^2}{4(n+1)^2}\bigg]\Bigg),
\end{equation*}
which is zero due to the divergence of the harmonic series. Therefore, $\mathrm{Kh}(u,v)=0$ for all $v\neq0$ and $|u|\geqslant1$. At the same time, in the region $|u|<1$, the product of \eqref{36-18} may converge to a certain non-zero value, since the function $|u|^n/\sqrt{n+1}$ decreases quite rapidly with increasing index.

When moving to complex values, significant difficulties arise. Assume that $it=u\in i\mathbb{R}$, while $v\in\mathbb{R}\setminus\{0\}$. This option corresponds exactly to the situation in formula \eqref{36-19}. In this case, the product in \eqref{36-18} is formally split into two parts $\mathrm{J}_j\cdot\mathrm{I}_j$, even in index "$n$" and odd, where
\begin{equation*}
\mathrm{J}_j(t,v)=
\prod_{n=1}^{\lfloor j/2\rfloor}J_0\bigg(\frac{t^{2n}v\sqrt{s}}{\sqrt{2n+1}}\bigg),
\,\,\,
\mathrm{I}_j(t,v)=
\prod_{n=1}^{\lceil j/2\rceil}I_0\bigg(\frac{t^{2n-1}v\sqrt{s}}{\sqrt{2n}}\bigg).
\end{equation*}
The "floor" and "ceiling" functions were used here. In this case, the function $\mathrm{J}_j(t,v)$ allows a similar analysis proposed above. However, there are difficulties with the function $\mathrm{I}_j(t,v)$. Indeed, let $|t|>1$ be fulfilled in addition to the above mentioned properties, then the product diverges, that is, $\mathrm{I}_j(t,v)\to+\infty$ for $j\to+\infty$. This is easy to verify, given the facts that $I_0(\cdot)$ is an even function strictly increasing on the positive part of axis, and the index function $t^{2n-1}/\sqrt{n}$ is bounded from below by a positive number for all $n\in\mathbb{N}$. In this case, there is such a $\delta_2>1$ that for all index values the relation is fulfilled
\begin{equation*}
I_0\bigg(\frac{t^{2n-1}v\sqrt{s}}{\sqrt{2n}}\bigg)\geqslant\delta_2.
\end{equation*}
Therefore, the divergence of the product follows from the estimate $\mathrm{I}_j(t,v)\geqslant\delta_2^{(j-1)/2}\to+\infty$ for $j\to+\infty$.
Similarly, using the divergence of the harmonic series, we can extend the result to the case of $|t|=1$. It is clear that in this case the limit function does not exist and it is not possible to permute the integral and the limit transition. It is necessary to investigate the functions $\mathrm{I}_j$ and $\mathrm{J}_j$ collectively, as the growth of one function can compensate for the decrease of the other. Thus, the limit transition in formula \eqref{36-18-1} is an auxiliary regularization that can clarify the meaning of the integral of the "bad" function. At the moment, only numerical analysis is possible for this area, and there are no rigorous mathematical proofs yet.

In the region $|t|<1$ and $v\in\mathbb{R}$, the function is finite. For the proof, we use the representation for the modified Bessel function of the first kind, see Section 8.431 in \cite{GR},
\begin{equation*}
I_0(t)=\frac{1}{\pi}\int_0^\pi\mathrm{d}s\,\mathrm{cosh}(t\,\mathrm{cos}(s))\leqslant e^{|t|}.
\end{equation*}
Then from the estimates
\begin{equation*}
\big|\mathrm{J}_j(t,v)\big|\leqslant1,
\,\,\,
\mathrm{I}_j(t,v)\leqslant\exp\Bigg(v\sqrt{s}\sum_{n=1}^{\lceil j/2\rceil}\frac{|t|^{2n-1}}{\sqrt{2n}}\Bigg),
\end{equation*}
it follows that the product is convergent and integrable with the weight $\exp(-t)$. Thus, the limit function $\mathrm{Kh}(it,v)$ exists and is finite.

Returning to formula \eqref{36-19}, we can say that the functional $\Psi$ can take finite values if the absolute values of the eigenvalues are "locked" in the box $|\lambda^a(x)\gamma|<1$. In all other cases, it is necessary to conduct additional research on the function $\mathrm{Kh}(\cdot\,,\cdot)$. Even if at some point the function tends to infinity (has a singularity), this does not mean that the functional $\Psi$ will tend to infinity, since formula \eqref{36-19} contains the integration operator $\mathrm{Tr}_{x,k}$, which can smooth out the behavior.

\section{Conclusion}
In this paper, an example of summation of a special kind of non-logarithmic singularities in all loops was presented. A new explicit formula was obtained, which, when formally decomposed by the coupling constant $\gamma$, leads to the sum of the power law terms with respect to the regularization parameter $\Lambda$ arising in the renormalized quantum action.

Let us pay attention to two interesting open questions. First, when setting the problem, a special class of non-logarithmic singularities was identified. Accordingly, summing the remaining parts proportional to $\Lambda^2$ is currently an unsolved problem. And although such additives will be included with a higher degree of the coupling constant, it is unclear how they will complement the existing contribution in \eqref{36-16}. Secondly, an interesting mathematical question is the study of properties of the function \eqref{36-18}. Especially in the region $i\mathbb{R}\times\mathbb{R}$, because it contains the data for the case of skew-symmetric matrices.

\vspace{2mm}
\noindent\textbf{Acknowledgements.} The work is supported by the Russian Science Foundation, grant 23-11-00311. The author expresses gratitude to N.V.Kharuk for useful comments, as well as to P.V.Akacevich and I.V.Korenev for fruitful joint work on a similar topic.

\end{document}